\documentclass[aps,prl,twocolumn,showpacs,superscriptaddress]{revtex4}
\usepackage{times}
\usepackage{epsfig}
\usepackage{graphicx}
\usepackage{amsmath}
\begin{document}

\title{Estimation of the Local Density of States on a Quantum Computer}

\author{Joseph Emerson}
\email[Corresponding author: ]{jemerson@mit.edu}
\affiliation{Department of Nuclear Engineering,  
Massachusetts Institute of Technology, Cambridge, MA, 02139}
\author{Seth Lloyd} 
\affiliation{Department of Mechanical Engineering, 
Massachusetts Institute of Technology, Cambridge, MA, 02139} 
\author{David Poulin} 
\affiliation{Institute for Quantum Computing, 
University of Waterloo, Waterloo, ON, N2L 3G1}
\author{David Cory}
\affiliation{Department of Nuclear Engineering, 
Massachusetts Institute of Technology, Cambridge, MA, 02139}

\date{\today}

\begin{abstract}
  We report an efficient
  quantum algorithm for estimating the local density of states (LDOS) 
  on a quantum computer.
  The LDOS describes the redistribution of
  energy levels of a quantum system under the influence of a 
  perturbation.  Sometimes known as the ``strength function'' from
  nuclear spectroscopy experiments, the shape of the LDOS is directly
  related to the survivial probability of unperturbed eigenstates, and
  has recently been related to the fidelity decay (or ``Loschmidt
  echo'') under imperfect motion-reversal.  
  For quantum systems that can be simulated 
  efficiently on a quantum computer, the LDOS 
  estimation algorithm enables an exponential speed-up over 
  direct classical computation.  
\end{abstract}


\maketitle

A major motivation for the physical realization of quantum information
processing is the idea, intimated by Feynman, that the dynamics of a
wide class of complex quantum systems may be simulated efficiently by
these techniques \cite{Lloyd}.  For a quantum system with Hilbert
space size $N$, an {\it efficient} simulation is one that requires only
Polylog($N$) gates.  This situation should be contrasted with direct
simulation on a classical processor, which requires resources growing
at least as $N^2$.  However, complete measurement of the final state
on a quantum processor requires $O(N^2)$ repetitions of the quantum
simulation.  Similarly, estimation of the eigenvalue spectrum of a
quantum system admitting a Polylog($N$) circuit decomposition requires
a phase-estimation circuit that grows as $O(N)$ \cite{AL97}.  As a
result there still remains the important problem of devising methods
for the efficient readout of those characteristic properties that are
of practical interest in the study of complex quantum systems.  In
this Letter we introduce an efficient quantum algorithm for
estimating, to $1/\mathrm{Polylog}(N)$ accuracy, the local density of
states (LDOS), a quantity of central interest in the description of
both many-body and complex few-body systems. We also determine
the class of physical problems for which 
the LDOS estimation algorithm provides an exponential speed-up over 
known classical algorithms given this finite accuracy.

The LDOS describes the profile of an eigenstate of an unperturbed
quantum system over the eigenbasis of perturbed version of the same
quantum system.  In the context of many-body systems the LDOS was
introduced to describe the effect of strong two-particle interactions
on the single particle (or single hole) eigenstates
\cite{Wigner,BohrMott,FGGK94,GS97,FI00}.  More recently, the LDOS has
been studied to characterize the effect of imperfections (due to
residual interactions between the qubits) in the operation of
quantum computers \cite{GS00,BCMS02}.  This profile plays
a fundamental role also in the analysis of system stability for
few-body systems subject to a sudden perturbation \cite{Heller}, such
as the onset of an external field, and has been studied extensively in
the context of quantum chaos and dynamical localization
\cite{BGI98,Izrailev}.  Quite generally the LDOS is related to the
survival probability of the unperturbed eigenstate
\cite{Heller,BohrMott,Jacquod01}, and there has been considerable
recent effort to understand the conditions under which the LDOS width
determines the rate of fidelity decay under imperfect motion-reversal
(``Loschmidt echo'') \cite{Jacquod01,Emerson02,Cucchietti01,WC01}.

A number of theoretical methods have been devised to characterize the
LDOS for complex systems. These methods include banded random
matrix models \cite{Wigner,FM95,Jacquod95,FCIC96}, models 
of a single-level with constant couplings to a ``picket-fence'' spectrum
\cite{BohrMott,Mellow}, and perturbative techniques
with partial summations over diagrams to infinite order \cite{CT}.  
Under inequivalent assumptions these approaches affirm a 
generic Breit-Wigner shape for the LDOS profile,
\begin{equation}
\eta^{\mathrm BW}(\phi) \propto \frac{\Gamma}{\phi^2 + \Gamma^2/4}.
\label{hyp1}
\end{equation}
However, the extent to which these methods correctly describe any 
real system is generally not clear \cite{Heller,Wisniacki02}, 
and therefore direct numerical analysis is usually 
necessary.  It is worth stressing here that direct  
numerical computation of the LDOS requires the diagonalization of 
matrices of dimension $N$,
and therefore demands resources that grow at least as $N^2$.
Of course only coarse-grained information about the LDOS is of
practical interest since one cannot even {\em store} the complete
information efficiently for large enough systems.
However, for generic systems there is no known numerical procedure that 
can circumvent the need to {\em manipulate} the $N\times N$ matrix 
in order to extract even coarse information about its LDOS. 
In this Letter we report a quantum algorithm
which enables estimation of the LDOS to 1/Polylog($N$) accuracy 
with only Polylog($N$) resources.

To specify the algorithm we represent the unperturbed quantum system
by a unitary operator $U$, which may correspond either to a Floquet
map, or to evolution under a time-independent Hamiltonian,
\begin{equation}
U  =   \exp(-i H_o \tau ) 
\end{equation}
We represent the perturbed quantum system by the unitary operator
$U(\sigma)$, which we express in the form,
\begin{equation}
U(\sigma)  =   \exp(-i \delta V ) U , 
\end{equation}
where $\delta$ is some dimensionless parameter and $V$ is a Hermitian
perturbation operator.  The variable $\sigma$ denotes an effective
``perturbation strength'' taking into account both the parameter $\delta$ 
and the size of the matrix elements of the perturbation,
\begin{equation}
\sigma^2 = 
\delta^2
\overline{|\langle \phi_j | V | \phi_{j'} \rangle|^2}
\end{equation}
where the average is taken only over directly coupled eigenstates.  
Let $U | \phi_j \rangle =
\exp(-i \phi_j) | \phi_j \rangle $, and $U(\sigma) | \phi_k(\sigma)
\rangle = \exp(- i \phi_k(\sigma)) | \phi_k(\sigma) \rangle$ denote
the eigenphases and eigenstates of the unperturbed and perturbed
systems respectively.  The LDOS for the $j$'th eigenstate of $U$ is
then,
\begin{equation}
\eta_{j}(\phi)  = \sum_k P\left(\phi_k(\sigma)|\phi_j \right) 
\; \delta\left( \phi - (\phi_k(\sigma) - \phi_j) \right), 
\end{equation}
where the transition probabilities,
\begin{equation}
P(\phi_k(\sigma)|\phi_j) = | \langle \phi_k(\sigma) | \phi_j \rangle  |^2,
\label{eqn:LDOS}
\end{equation}
are the basic quantities of interest. 

The coarse-grained distribution,
\begin{equation}
P(\Delta_l | \phi_j) =  \sum_{\phi_k(\sigma) \in \Delta_l} 
P(\phi_k(\sigma)|\phi_j), 
\label{eqn:cLDOS}
\end{equation}
is a just the sum over the probabilities for those perturbed
eigenphases $\phi_k(\sigma)$ lying within a band $\Delta_l$.  This
band is centered about angle $2\pi l /M$, with width $\Delta = 2
\pi/M$, and the integer $l$ ranges from $0$ to $M-1$.  Similarly, an
averaging over neighboring {\em unperturbed} eigenstates is often
carried out to remove the effects of atypical states. 
The combination of both
operations yields the probability distribution,
\begin{equation}
P(\Delta_l | \Delta_m) = N_m^{-1}  
\sum_{\phi_k(\sigma) \in \Delta_l} 
\sum_{\phi_j \in \Delta_m} 
P(\phi_k(\sigma)|\phi_j), 
\label{eqn:acLDOS}
\end{equation}
where the normalization constant $N_m$ is just the number of
unperturbed eigenphases in the angular range $\Delta_m$. 
In practice one must choose $M$ to be $O(\log(N))$ since otherwise the
measured LDOS $\eta$ would contain an exponential amount of
information and therefore could not be processed efficiently.

\begin{figure}[tbh]
\centering
\includegraphics*[width = 8.7cm]{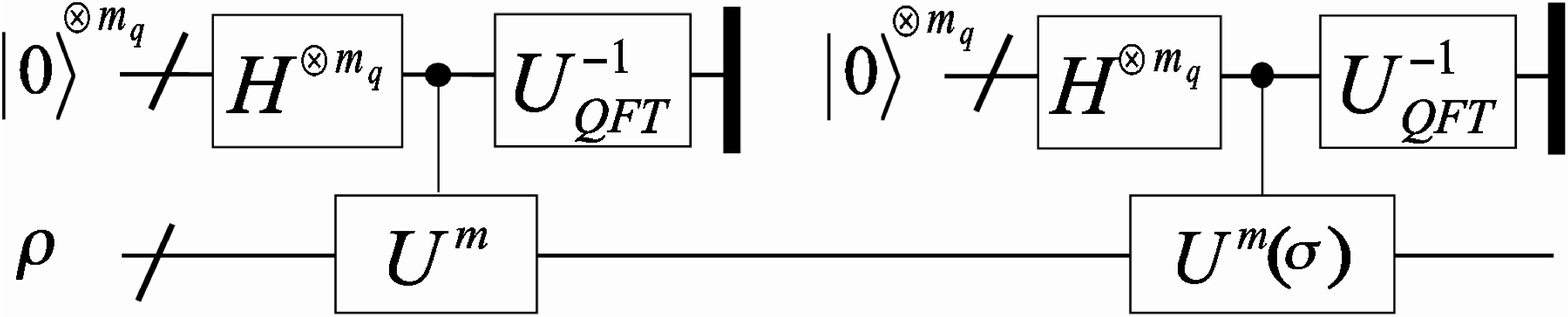}
\caption{
  Circuit diagram for measuring the local density of states,
  consisting of two successive phase-estimation circuits on different
  operators.  The diagonal line denotes a bundle of qubits and the
  thick vertical bar denotes a projective measurement of the quantum
  state in the computational basis. The upper register contains $m_q =
  \log(M)$ qubits and the operations on the lower register are applied
  conditionally $m$ times, where the integer $m \in [0, M-1]$ is
  determined from the binary representation of the computational basis
  states in the upper register.}
\label{circuit}
\end{figure}

We now describe the algorithm for estimating the LDOS on a quantum
processor.  The circuit for this algorithm is depicted in Fig.~1.  The
lower register implements the perturbed and unperturbed maps $U$
and $U(\sigma)$, requiring $n_q  = O(\log_2(N))$ qubits.  The upper
register holds the $m_q = \log_2(M)$ ancillary qubits which fix the
precision of the phase-estimation algorithm.  The upper register
always starts out in the `ready' state $|0\rangle$.  The appropriate
choice of initial state $\rho$ in the $n_q$ register will depend on
the context, as explained below.  For the moment we assume the lower
register is prepared in a pure state, $\rho = | \psi_o \rangle \langle
\psi_o |$.  The first step of the algorithm involves estimating the
eigenphases of the unperturbed operator $U$.  This takes the initial
state through the sequence,
\begin{eqnarray}
 |0\rangle \otimes | \psi_o \rangle 
& \rightarrow &\frac{1}{\sqrt M}\sum_{m=0}^{M-1} 
        | m \rangle | \psi_o \rangle \nonumber \\
&\rightarrow & 
        \frac{1}{\sqrt{M} }\sum_{m=0}^{M-1} 
| m \rangle (U)^m | \psi_o \rangle \nonumber \\ 
 & = & \frac{1}{\sqrt M }\sum_{m=0}^{M-1} 
| m \rangle
\sum_{j=0}^{N-1} c_j \exp( i \phi_j m) | \phi_j \rangle \nonumber \\ 
 &\rightarrow & 
\sum_{j=0}^{N-1} c_j | m_j
\rangle  | \phi_j \rangle,  
\end{eqnarray}
where $c_j = \langle \phi_j | \psi_o \rangle$. The state $m_j$ is the
nearest $m_q$-bit binary approximation to the $j$'th eigenphase of
$U$
\begin{equation}
{\tilde \phi_j} = 2 \pi m_j /M \simeq \phi_j. 
\end{equation}
Upon strong measurement of the $m_q$ register one obtains and records
a single outcome $m$, and the state of $n_q$ register must then be
described by (viz, `collapsed to') the updated pure state,
\begin{equation}
| \psi(\Delta_m)  \rangle 
=  \sum_{\phi_j \in \Delta_m} \tilde{c}_j | \phi_j \rangle,  
\end{equation}
corresponding to the subspace of eigenstates with eigenphases 
in the band $\Delta_m$ of width  
$\Delta = 2 \pi/ M$ about the phase $2 \pi m /M$. 
To keep normalization 
the coefficients have been rescaled as follows,   
\begin{equation}
\tilde{c}_j = \frac{c_j}{ ( \sum_{\phi_j \in \Delta_m} |c_j|^2 )^{1/2}} 
\end{equation}
Next we reset the $m_q$ qubit register to the ready state and run the
phase-estimation algorithm on the operator $U(\sigma)$, producing the
final state,
\begin{equation}
 | \psi \rangle  = 
 \sum_{\phi_j \in \Delta_m} \tilde{c}_j 
\sum_{k=0}^{N-1} b(k|j) \;  
| m_k
\rangle \otimes | \phi_k(\sigma) \rangle, 
\end{equation}
where ${\tilde \phi_k(l)} = 2 \pi m_k /M $ is an $m_q$-bit
approximation to $\phi_k(\sigma)$. The complex coefficients $b(k|j) =
\langle \phi_j |\phi_k(\sigma) \rangle $ are the inner product of the
perturbed and unperturbed eigenstates.  Measurement of the $m_q$
register now reveals an outcome $l$, associated with the eigenphases
in the angular range $2\pi l/M \pm \Delta/2$.  The outcome $l$ occurs
with probability,
\begin{equation}
P_{\psi_o}(l|m) = \sum_{\phi_k \in \Delta_l} 
\left| \sum_{\phi_j \in \Delta_m}  \tilde{c}_j \; b(k|j) \right|^2 
\label{genP}
\end{equation}
which is conditional on the earlier outcome $m$ and 
the choice of initial state. 

We now specify how the initial state may be chosen to eliminate
unwanted fluctuations arising from the variables $\tilde{c}_j$ in
Eq.~\ref{genP} .  Before describing the general solution we consider
first a special case of particular interest: when a known eigenstate
of $U$ may be prepared efficiently.  Such an initial state may be
prepared (or well approximated) by an efficient circuit when $U$
consists of some sufficiently simple integrable system (e.g., a
non-interacting many-body system).  In this case we have $ \tilde{c}_j
= \delta_{jk}$, and the final probability
distribution Eq.~\ref{genP} reduces exactly to the (coarse-grained)
kernel Eq.~\ref{eqn:cLDOS},
\begin{equation}
P_{\phi_k}(l|m) \rightarrow P(\Delta_l| \phi_k). 
\end{equation}
When the eigenphase associated to the prepared eigenstate is
known to sufficient accuracy (so that $m$ is known), 
it is not even necessary to perform the first
phase estimation routine.  
In the general case of a generic quantum system, it is sufficient to 
prepare the maximally mixed state as the initial state, in which case the final probability
distribution reduces exactly to the (coarse-grained and averaged) 
probability kernel Eq.~\ref{eqn:acLDOS}, i.e., 
\begin{equation}
        P_{{\bf 1} /N}(l|m) =  \frac{1}{N_m} 
\sum_{\phi_k(\sigma) \in \Delta_l} 
\sum_{\phi_j \in \Delta_m} 
P(\phi_k(\sigma)|\phi_j). 
\end{equation}
This probability kernel contains all the information needed to compute the
(coarse-grained and averaged) LDOS, 
$\eta_m(2\pi k/M) = \sum_l P(l|m) \delta_{k,(l-m)}$, completing our derivation.

The algorithm described above remains efficient provided that 
the quantum maps $U$ and $U_\delta$
admit Polylog($N$) gate decompositions.  Such decompositions have
been identified both for many-body systems with local interactions and
for a wide class of few-body quantized classical models. As mentioned
earlier, for practical purpose $M$ should be Polylog($N$) so the
overall circuit of Fig.~\ref{circuit} is indeed efficient for such
systems.

We now turn to the question of how many times $K$ the algorithm 
must be repeated to arrive at interesting physical conclusions 
about the final probability distribution. 
This issue arise because the final probability distribution is not measured 
directly on the quantum processor; rather, it 
governs the relative frequency of outcomes obtained in each 
repetition of the algorithm. Indeed, it is 
by repeating the algorithm illustrated in Fig.~\ref{circuit} and
accumulating joint statistics of the $l$ and $m$ outputs that one 
can estimate the parent distribution $P(l|m)$. The
accuracy of this estimation depends on the number of times $K$ the
distribution is sampled. In order to bound $K$ it is convenient to cast 
the physical problems related to the LDOS in terms of hypothesis testing.  
We consider the important case of testing which of two candidates 
distributions $\eta_1$ or $\eta_2$ best describes the LDOS of 
a given system and a given perturbation. For example, 
one might be testing whether the Lorentzian has one of two 
candidate widths, or whether the profile is Gaussian or Lorentzian. 
Only when $K\leq \mathrm{Polylog}(N)$ will the overall computation 
remain efficient.
This problem is resolved in general by the Chernoff bound \cite{CT1991}. 
A random variable is distributed according to either $P_1(x)$ or $P_2(x)$, and
we wish to determine which distribution is the right one. Then, the
probability $P_e$ that we make an incorrect inference decreases exponentially 
with the number of times $K$ the variable was sampled: $P_e \leq \lambda^K$.
Here, $0\leq\lambda\leq 1$ is a measure of similarity between
distributions defined as
\begin{equation}
\lambda = \min_{0\leq\alpha\leq 1} \sum_x P_1(x)^\alpha P_2(x)^{(1-\alpha)};
\end{equation}
in particular, $\lambda$ is bounded above by the fidelity between
$P_1$ and $P_2$. Thus, a constant error probability $\epsilon$
requires a sample of size $K = \log(\epsilon)/\log(\lambda)$.
Therefore, as long as the concerned distributions are at a
Polylog($N$) distance, i.e.  $1-\lambda \geq 1/\mathrm{Polylog}(N)$,
they can be distinguished efficiently. We note that the test can be
inconclusive when both hypothesis are equally likely to describe the
underlying physics. 

Efficient application of the LDOS algorithm under these restrictions
may be illustrated explicitly by working through a problem of
practical interest from the recent literature.  We consider the
problem of testing whether the Breit-Wigner profile Eq.~\ref{hyp1}
applies when a given quantized classically chaotic model is subjected
to a perturbation of interest.  From the BGS conjecture
\cite{BT77BGS84} and studies of (banded) random matrix models
\cite{Wigner,FM95,Jacquod95,FCIC96}, it is generally expected that for
fully chaotic models with generic perturbations Eq.~\ref{hyp1} applies
with,
\begin{equation}
\Gamma(\sigma) = 2\pi \sigma^2 \rho_E,  
\label{gamma}
\end{equation}
provided that the effective perturbation strength lies in the range,
\begin{equation}
1 \ll \sigma \rho_E  \ll \sqrt{b}, 
\label{hyp2}
\end{equation}
where $b$ is the bandwidth of the perturbation in the ordered 
eigenbasis of $U$ and $\rho_E$ is the level density. 
It should be stressed that $\Gamma$ may be estimated a priori 
if the perturbation is known \cite{Emerson02,FM95,Jacquod95}.  
Deviations from this hypothesis can arise for a wide variety 
of reasons (i.e., integrable or mixed classical dynamics in the unperturbed or 
perturbed system, non-generic properties of the perturbation, hidden 
symmetries, etc) and therefore analysis of the LDOS remains 
an active area of numerical study for both dynamical models \cite{BGI98,Wisniacki02} 
and real systems \cite{FGGK94}. 

The lower bound of Eq.~\ref{hyp2} is determined from the breakdown 
of perturbation theory and leads to a width $\Gamma$ that decreases 
linearly with $N$.  
Since the circuit can only efficiently resolve the LDOS with 
accuracy 1/Polylog($N$), 
the BW profile with width $\Gamma = O(N^{-1})$ may not be verified 
efficiently near this lower bound.  However, 
near the upper bound of Eq.~\ref{hyp2} the validity of the BW profile 
may be tested efficiently.  
In the case of fully chaotic models one has $b = N/2$ and the upper 
bound for $\Gamma$ is therefore $O(1)$.
Hence the validity of Eq.~\ref{hyp1} provides a hypothesis which may 
be tested efficiently for any perturbation  
such that $ 1/\mathrm{Polylog}(N) \ll \Gamma(\sigma) \ll O(1)$.
Near this bound one can also determine whether the chaotic model exhibits 
dynamical localization, since in this case one has a bandwidth $b \ll N/2$ 
and the LDOS will cease to maintain 
the BW profile when $ b \rho_E^{-1} \ll \Gamma(\sigma) \ll O(1)$.
Indeed for some models the localization length $l$ of the eigenstates scales 
as $l \simeq O(1)$ \cite{Simone}, and hence this length may be estimated 
using the LDOS algorithm with only Polylog($N$) resources. 

In summary we have reported an algorithm for efficiently 
estimating the LDOS of a quantum system subject to perturbation. 
There is wide range of contexts in 
which important coarse features of the LDOS, such as the width, 
may be estimated with only Polylog($N$) resources. 
We have described in detail the important problem of testing 
the Breit-Wigner hypothesis as one example for which the LDOS 
estimation algorithm 
gives an effective exponential speed up over classical computation. 

We are grateful to D. Shepelyanksy and Y. Weinstein for
helpful discussions.  This work was supported by the NSF and CMI.

\end{document}